\documentclass[twocolumn,prb,showpacs]{revtex4}
\usepackage[latin9]{inputenc}
\usepackage{amsmath}
\usepackage{amssymb}
\usepackage{graphicx}
\usepackage{esint}
\usepackage[dvipdfm,
            pdfstartview=FitH,
            CJKbookmarks=true,
            bookmarksnumbered=true,
            bookmarksopen=true,
            colorlinks,
            pdfborder=001,
            linkcolor=blue,
            anchorcolor=blue,
            citecolor=blue
            ]{hyperref}
\usepackage{longtable}

\makeatletter
\@ifundefined{textcolor}{}
{%
 \definecolor{BLACK}{gray}{0}
 \definecolor{WHITE}{gray}{1}
 \definecolor{RED}{rgb}{1,0,0}
 \definecolor{GREEN}{rgb}{0,1,0}
 \definecolor{BLUE}{rgb}{0,0,1}
 \definecolor{CYAN}{cmyk}{1,0,0,0}
 \definecolor{MAGENTA}{cmyk}{0,1,0,0}
 \definecolor{YELLOW}{cmyk}{0,0,1,0}
}

\usepackage{stmaryrd}\usepackage{dcolumn}\usepackage{bm}\usepackage{CJK}

\makeatother

\begin{document}


\begin{titlepage}

\title{Defect energetics and magnetic properties of 3$d$-transition-metal-doped topological crystalline insulator SnTe}

\author{Na \surname{Wang}$^{1}$}
\author{Jianfeng \surname{Wang}$^{1}$}
\author{Chen \surname{Si}$^{2}$}
\author{Bing-lin \surname{Gu}$^{1,3,4}$}
\author{Wenhui \surname{Duan}$^{1,3,4}$}\email{dwh@phys.tsinghua.edu.cn}

\affiliation{
$^{1}$Department of Physics and State Key Laboratory of Low-Dimensional Quantum Physics, Tsinghua University, Beijing 100084, People's Republic of China\\
$^{2}$School of Materials Science and Engineering, Beihang University, Beijing 100191, People's Republic of China\\
$^{3}$Institute for Advanced Study, Tsinghua University, Beijing 100084, People's Republic of China\\
$^{4}$Collaborative Innovation Center of Quantum Matter, Tsinghua University, Beijing 100084, People's Republic of China
}

\date{\today}

\begin{abstract}
The introduction of magnetism in SnTe-class topological crystalline insulators is a challenging subject with great importance in the quantum device applications. Based on the first-principles calculations, we have studied the defect energetics and magnetic properties of 3$d$ transition-metal (TM)-doped SnTe. We find that the doped TM atoms prefer to stay in the neutral states and have comparatively high formation energies, suggesting that the uniform TM doping in SnTe with a higher concentration will be difficult unless clustering. In the dilute doping regime, all the magnetic TM atoms are in the high-spin states, indicating that the spin splitting energy of 3$d$ TM is stronger than the crystal splitting energy of the SnTe ligand. Importantly, Mn-doped SnTe has relatively low defect formation energy, largest local magnetic moment, and no defect levels in the bulk gap, suggesting that Mn is a promising magnetic dopant to realize the magnetic order for the theoretically-proposed large-Chern-number quantum anomalous Hall effect (QAHE) in SnTe.
\end{abstract}


\draft

\vspace{2mm}

\maketitle

\end{titlepage}

\section{Introduction}

Topological crystalline insulator (TCI) \cite{TCI,sntefu}, a recently discovered new class of symmetry-protected topological states with an insulating bulk gap and gapless surface states, has triggered much interest in both the modern theoretical physics and the technological applications. Protected by the crystal symmetry, TCIs have multiple branches of Dirac-like surface states on the surfaces with the underlying symmetry. The TCI phase has been experimentally realized in SnTe-class IV-VI semiconductors \cite{snpbse_exp,snpbte_exp,snte_exp}, and various of exotic properties and phenomena have been studied, such as electrically tunable spin-filtered edge states in SnTe thin film, topological superconductivity in indium doped SnTe, strain-tunable valleytronics, and so on \cite{Junwei_natmater_2014,Majorana,TCI-vallylu,TCI_vally,Junwei_prb_2013,
OkadaYoshinori_science_2013_gap-TCI,GangYang_prb_2014,FangChen_prl_2014,
Junwei_nano_2015}. Especially, when an out-of-plane ferromagnetic order is introduced into a TCI film, the QAHE with a tunable large Chern number can be produced \cite{NiuChengwang_prb_2015_2d-TCI-QAHE,FangChen_prl_2014_QAHE}, distinguished from the case of $\pm$ 1/$\pm$ 2 Chern-number QAHE in topological insulator (TI) films \cite{YuRui_science_2010,cuizu_QAHE,TI_QAHE_Bi}. This has motivated many efforts to introduce magnetism in TCIs by magnetic doping or proximity effects with magnetic substrates \cite{Assaf_prb_2015_Eu-SnTe,Kilanski_arxiv_2013_Cr-SnTe,Fe-SnTe}.

As an important and controllable method, magnetic doping with 3$d$ TM atoms has been successfully applied in Bi$_{\rm 2}$Te$_{\rm 3}$-class TIs \cite{Jinsong_science_2013,Cuizu_prl_2014,cuizu_QAHE}, and QAHE has been realized in experiments in Cr-doped (Bi$_{x}$Sb$_{{\rm 1-}x}$)$_{\rm 2}$Te$_{\rm 3}$ films\cite{YuRui_science_2010,cuizu_QAHE,TI_QAHE_Bi}. In the past few years, some experimental efforts have also been devoted to study the magnetism in SnTe doped with 3$d$ TMs \cite{InoueMasasi_jlowtempphys_1976_MnSnTe,W.J.M.de-Jonge_semicondscitechnol_1990_carrier-MnSnTe,InoueMasasi_1981}. It is found that SnTe shows complex magnetic properties under different TM doping: for example, it displays ferromagnetic behavior under the doping of Mn, Cr or Fe, while it is magnetically ineffective under doping of Co or Ni \cite{InoueMasasi_jlowtempphys_1976_MnSnTe,InoueMasasi_1981}. However, the underlying physical mechanisms for these different magnetic behaviors are not clear yet. Moreover, effective magnetic doping in SnTe is still a challenge --- well-defined controllable magnetism in SnTe has never been achieved by TM doping, preventing the further investigations about magnetic effect in TCIs.
Recently, the electronic and magnetic properties of V, Cr or Mn-doped SnTe are theoretically studied\cite{LiuY_jap_2013,LiuY_JMMM_2015}. However, the high doping concentration considered in these studies doesn't agree with the realistic experiment \cite{InoueMasasi_jlowtempphys_1976_MnSnTe,W.J.M.de-Jonge_semicondscitechnol_1990_carrier-MnSnTe,InoueMasasi_1981}, and probably suppresses the TCI phase. Therefore, it is highly desirable to investigate the electronic properties of TM-doped SnTe with a realistic experimental doping concentration and find the underlying mechanism for the arising of magnetism.

In this work, we investigated the defect energetics and the local magnetic states of the 3$d$ TMs dilutely doped in SnTe by using the first-principles calculations, aiming to examine whether the magnetic 3$d$ TM atoms can be effectively doped in SnTe and study the magnetic properties of SnTe under the realistic experimental doping concentration. We find that 3$d$-TM-doped SnTe has relatively high formation energy, which is positively correlated with the cohesive energy of the elemental TM bulk, suggesting that the uniform TM doping in SnTe with a higher concentration will be difficult. The TMs prefer to stay in the neutral states when the Fermi level is located in the gap or below the valence band maximum (VBM) as in realistic SnTe. All the magnetic atoms show high-spin electronic configurations, indicating that the spin splitting energy of 3$d$ TM is stronger than the crystal splitting energy of the SnTe ligand. Without regard to magnetic coupling between the TM atoms in the dilute doping regime, the TM atoms in SnTe show nearly the same magnetic moments as the isolated TM atoms, except Sc and Cr. Moreover, with relative low doping formation energy, Mn-doped SnTe possesses the largest magnetic moments and no defect levels in the bulk gap, serving as a prominent candidate for the magnetic investigation in SnTe.

\begin{figure}[tbp]
\includegraphics[width=0.4\textwidth]{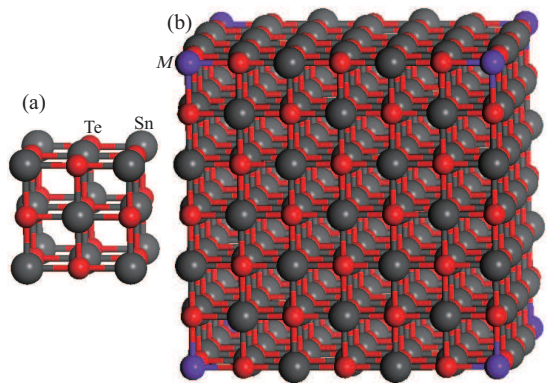}
\caption{\label{fig:structure}(Color online) (a) Structures of SnTe in the conventional cell of the rock-salt structure and (b) substitutional defect $M_{\rm Sn}$ shown in a 3$\times$3$\times$3 supercell. Sn and Te are represented by large-gray and small-red spheres, respectively. The TM atom is represented by violet sphere in (b), labeled as $M$.}
\end{figure}

\section{methods}
All the calculations are based on the density functional theory within the Perdew-Burke-Ernzerhof parameterization of the generalized gradient approximation (GGA) \cite{GGA}, as implemented in the Vienna ab initio simulation package \cite{VASP}. Interactions between ion cores and valence electrons are described by the projector augmented wave \cite{PAW} method. Plane waves with a kinetic energy cutoff of 250 eV are used as the basis set. Here only the substitutional defect [$M_{\rm Sn}$, with $M$ representing the 3$d$ TM elements (Sc--Zn)] is considered as the doping configuration of the TM in SnTe, as the dominant intrinsic defect of SnTe is the Sn vacancy, and the antisite and interstitial defects are highly energetic unfavored \cite{snte_exp,mine}. Moreover, the previous experiments also showed that the TM atoms will substitute the Sn ions, such as Sn$_{1-x}$Mn$_x$Te \cite{InoueMasasi_1981,SYeohT_JPSJ_1988_strucMSnTe}. To avoid the interaction between image TM atoms in the neighbouring supercells, a 3~$\times$~3~$\times$~3 supercell consisting of 216 atoms is used to construct the isolated defect structure, as is shown in Fig. \ref{fig:structure}, which corresponds to a low doping level of 0.4\%. The integration over the Brillouin zone is done with 3~$\times$~3~$\times$~3 grid points \cite{MP}. All of the structures are fully relaxed within the spin-polarized calculation until the maximum residual ionic force is below 0.01 eV/\AA. Following the treatment of the 3$d$ TM-doped Bi$_{\rm 2}$Te$_{\rm 3}$-class TIs \cite{ZhangJianmin_PRB_2013_M-TI}, the Hubbard effect U of the TM atoms is not taken into account. In fact, we have done test calculations, and find that U only affects the cohesive energy of the elemental TM bulk and the doping formation energy will be pushed to higher. Moreover, U hardly influences the local magnetic states. We also find that spin-orbit coupling (SOC) rarely influences the formation energy and the occupation of the $d$ orbitals, so the following discussion will be mainly based on the spin-polarized calculations.

The defect energetics are determined by the calculations of formation energy of defects, which is defined as \cite{mine,chrisriview}
\begin{eqnarray}
\Delta H [{M}_{\rm Sn}^{q}] & = & E_{\rm tot}[{M}_{\rm Sn}^{q}]-E_{\rm tot}[{\rm bulk}]+\mu_{\rm Sn}-\mu_{M}\nonumber \\
 &  & +q(E_{F}+E_{v}+\Delta V), \label{formation}
\end{eqnarray}
where $E_{\rm tot}$ [$M_{\rm Sn}^q$] and $E_{\rm tot}$ [bulk] are the total energies of a supercell with and without substitutional defect $M_{\rm Sn}$, respectively, and $\mu_{\rm Sn}$ and $\mu_{M}$ represent the chemical potentials of Sn and the TM atoms. In experiments, the Sn-rich condition is usually used for SnTe growth to reduce the Sn vacancies, so here we also choose the Sn-rich condition, i.e., $\mu_{\rm Sn}$ is set to be the energy of diamond cubic phase of bulk Sn \cite{cryst.struct.}. Meanwhile, $\mu_{M}$ is also chosen to be the energy of the bulk phase of the magnetic element, representing the case of plenty of magnetic source for doping. The reference magnetic elemental bulk phases and their magnetization states are all taken from Ref. \onlinecite{CRC}. The last term in Eq. (1) indicates the charged-defect-state-dependent energy, in which $q$ is the charge on the defect and $E_F$ is the Fermi level referenced to the VBM in the bulk. Due to the choice of this reference for $E_F$, we need to explicitly insert the energy of the bulk VBM, $E_v$, in our expressions for formation energies of charged states. $\Delta V$ is to align the reference potential in our defect supercell with that in the bulk \cite{mine}.

\begin{figure}[tbp]
\includegraphics[width=0.45\textwidth]{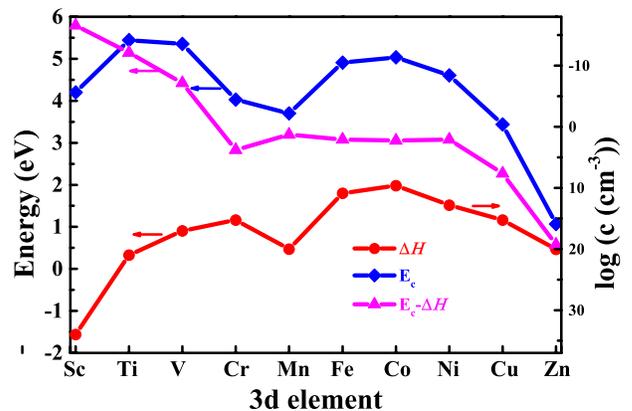}
\caption{\label{fig:form_n}(Color online) (a) The formation energies $\Delta H$ (red) of 3$d$ TM-doped SnTe, the cohesive energies $E_{c (M)}$ (blue) of the bulk phase of TM elements, and the difference $E_c$-$\Delta H$ (magenta) between the former two energies as functions of 3$d$ TM elements.
}
\end{figure}
\section{Results and Discussion}
First, let's concentrate on the neutral defect of 3$d$ TM doping ($q$ = 0). Figure 2 shows the calculated formation energies of substitutional defects ($M_{\rm Sn}$) for all 3$d$ TMs (Sc--Zn) in SnTe. It can be seen that the formation energy curve appears as a M shape, with a drop at Mn.
The formation energies for Sc, Ti, Mn, Zn are relatively low. Especially, the formation energy of Sc$_{\rm Sn}$ is negative, indicating that Sc doping may bring the instability of SnTe.
The formation energy of Mn is lowest among the middle TM atoms (V--Ni), which is consistent with the experimental result that the solubility of Mn in SnTe is largest and it is quite limited for other middle TMs \cite{InoueMasasi_1981,InoueMasasi_jlowtempphys_1976_MnSnTe,W.J.M.de-Jonge_semicondscitechnol_1990_carrier-MnSnTe,Mn-solubility,Fe-solubility,
Kilanski_arxiv_2013_Cr-SnTe,Cr-clustering-in-SnTe}. We also calculate the cohesive energy of the bulk phase of TM element ($E_c$) as shown in Fig. \ref{fig:form_n}, consistent with the experimental values \cite{cohe-3d-TM}. The drop of the cohesive energy at Mn can be understood as that the strong screening of the half-filled-$d$-orbital electrons
reduces the interatomic coupling in elemental Mn\cite{radius}. It can be found that, the M-shape trend of formation energy mentioned above is quite similar to that of the cohesive energy of the bulk phase of TM elements.
Interestingly, the difference between the cohesive energy $E_c$ and formation energy $\Delta H$ ($E_c-\Delta H$, indicated by the magenta line in Fig. \ref{fig:form_n} (a)) is nearly a constant from Cr to Ni. This reveals that these TM atoms have similar bonding with the host when doped in SnTe, which may be related to the same valence state (+2) of dopants in SnTe. However, for the early (Sc--V) and late (Cu--Zn) TM ions, $E_c-\Delta H$ varies significantly, indicating different bonding between the TM ion and the SnTe host (large atomic radius distinction exists between the early/late TMs and Sn \cite{CRC}).

According to the concentration formula c = $N_{\rm sites}$e$^{-\Delta H/kT}$, where $N_{\rm sites}$ is the site concentration, and $T$ is the temperature (chosen to be the usual experimental temperature 700 K for SnTe growth here), the doping concentrations of the TM under the thermoequilibrium condition can be estimated, as shown on the right axis of Fig. \ref{fig:form_n}. However, it can be found that the concentrations for the TMs doping in SnTe are comparatively low. The doping concentration can only reach 10$^{19-20}$ cm$^{-3}$ (0.01--0.1\%) even for Ti, Mn and Zn. This is completely different from the case of TM doping in Bi$_{\rm 2}$Te$_{\rm 3}$-class TIs --- V and Cr substitutions have low formation energy ensuring the possibility of spontaneous doping \cite{ZhangJianmin_PRB_2013_M-TI}. Instead, high-level doping of TM in SnTe will bring in TM elemental clustering in SnTe, as experimentally observed \cite{Kilanski_arxiv_2013_Cr-SnTe,Fe-solubility,Cr-clustering-in-SnTe}. The discrepancy for TM doping in SnTe and Bi$_{\rm 2}$Te$_{\rm 3}$ may originate from the distinction of their crystal structure: the strain introduced by the TM doping cannot be released in the rock-salt structure of SnTe \cite{sntefu,mine} but can easily be released in the van der Waals layered structure of Bi$_{\rm 2}$Te$_{\rm 3}$ \cite{haijun_3dTI,mine-surf}. This hard-doping problem of TM atoms widely exists in the covalent or ionic block materials, such as in Si, GaAs, MgO, etc \cite{Cr-cluster-in-GaN,TM-clustering-in-Si,TM-cluster-in-Ge,TM-clustering}.

\begin{figure}[tbp]
\includegraphics[width=0.4\textwidth]{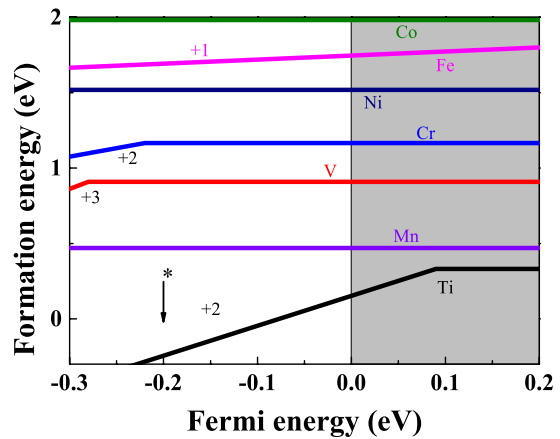}
\caption{\label{fig:form_c}(Color online) Formation energy as a function of the Fermi level for the 3$d$ TM-doped SnTe. Zero of the Fermi level is set to be the VBM of the host. The slope of each segment denotes the charged state of defect, which is labeled beside each segment. Kinks in the curves indicate thermodynamic transition levels between different charged states. The gray rectangle encloses the bulk band gap region of SnTe. The arrow marked by the star points the Fermi level of the intrinsic SnTe under Sn-rich condition as a result of Sn vacancy. In order for comparison, the formation energy for Mn, only under the neural state, is also given.}
\end{figure}

As the conductivity of SnTe is always $p$-type doped, induced by the negatively charged Sn vacancy in the realistic environment \cite{snte_exp,mine}, the TM dopants in SnTe may not stay in the neutral state, but become positively charged, i.e., the electrons localized around the TM ion may be lift into the bulk bands of SnTe. In this regard, we further investigate the charged state of the TMs (Ti--Cr, Fe--Ni) that have localized $d$ orbitals near the gap region by removing the electrons of these localized orbitals. In Fig. \ref{fig:form_c}, the charged states with the lowest energy of each defect are shown for any given $E_F$. In order for comparison, the formation energy for Mn, only under the neural state, is also given. The slope of the lines corresponds to the charge of defects, with kinks appearing at the thermodynamic transition levels between the different charged states of a given defect.
The VBM of the host is set to be zero of the Fermi level in Fig. \ref{fig:form_c}, and the shaded region spans the bulk gap of SnTe.
Due to the Sn vacancy \cite{mine}, the Fermi level of the intrinsic SnTe can be pushed down to around -0.2 eV below the VBM (pointed by the star in Fig. \ref{fig:form_c}).
It can be found that the TM dopants tend to keep neutral states in the gap region and even at the Fermi level down to -0.2 eV.
One exception is Fe dopant: Fe tends to stay in the charged state of +1, serving as dopant; however, its formation energy is extremely high which limits its doping. Ti prefers to be of +2 charged state with lower formation energy; nevertheless, Ti will lose its local magnetic moment when charged. From the above analysis, we can conclude that under the realistic situation only the neutral state of the TM dopant needs to be considered.

\begin{table*}
\caption {The electronic configurations (Config) and the magnetic moments ($\mu_{\rm atom}$) for the isolated 3$d$ TM atom \cite{book_chemphy} are listed in the first two rows. The calculated magnetic moments ($\mu$) for the 3$d$ TM atoms in SnTe are listed in the third row. The magnetic moments and the corresponding electronic configurations (in braket) for the 3$d$ TM ions with 4--7 $d$ electrons in the octahedral crystal field for the high-spin and low-spin states \cite{book_chemphy} are listed in the HS and LS rows. The predicted valence for 3$d$ TM atoms in SnTe is listed in the last row.} \label{tab:mu}
    \begin{tabular}{|l|l|l|l|l|l|l|l|l|l|l|}
    \hline
    Element               & Sc            & Ti           & V            & Cr           & Mn           & Fe           & Co           & Ni           & Cu            &Zn             \\ \hline
    Config                & 3$d^1$4$s^2$  & 3$d^2$4$s^2$ & 3$d^3$4$s^2$ & 3$d^5$4$s^1$ & 3$d^5$4$s^2$ & 3$d^6$4$s^2$ & 3$d^7$4$s^2$ & 3$d^8$4$s^2$ & 3$d^{10}$4$s^1$ & 3$d^{10}$4$s^2$  \\ \hline
    $\mu_{\rm atom}/\mu_B$& 1             & 2            & 3            & 5            & 5            & 4            & 3            & 2            & 0           & 0             \\ \hline
    $\mu/\mu_B$  & 0             & 1.56         & 3            & 4            & 5            & 4            & 2.80         & 1.87         & 0           & 0             \\ \hline
    HS$/\mu_B$       &- &- &- & 4 ($t_{2g}^{\uparrow,3}e_{g}^{\uparrow,1}$) & 5 ($t_{2g}^{\uparrow,3}e_{g}^{\uparrow,2}$) & 4 ($t_{2g}^{\uparrow,3;\downarrow,1}e_{g}^{\uparrow,2}$) & 3 ($t_{2g}^{\uparrow,3;\downarrow,2}e_{g}^{\uparrow,2}$) &- &- &- \\ \hline
    LS$/\mu_B$       &- &- &- & 2 ($t_{2g}^{\uparrow,3;\downarrow,1}$) & 1 ($t_{2g}^{\uparrow,3;\downarrow,2}$) & 0 ($t_{2g}^{\uparrow,3;\downarrow,3}$) & 1 ($t_{2g}^{\uparrow,3;\downarrow,3}e_{g}^{\uparrow,1}$) &- &- &- \\ \hline
    valence               & +3            & +2           & +2           & +2           & +2           & +2           & +2           & +2           & +1           & +2            \\ \hline
    \end{tabular}
\end{table*}


\begin{figure}[tbp]
\includegraphics[width=.4\textwidth]{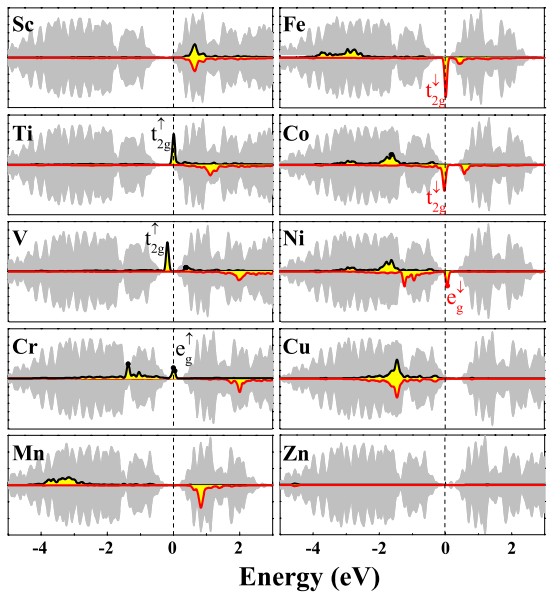}
\caption{\label{fig:dos}(Color online) The density of states (DOS) of TM-doped SnTe (gray-shaded) and the partial DOS of the $d$ orbitals of the single TM in SnTe (yellow-shaded, upper and lower parts for spin-up and spin-down states, respectively) within the spin-polarized calculation. For clearance, the partial DOS of TM is amplified 5 times. The Fermi level is set as zero in each panel, labeled by the dashed lines. The orbitals that could be clearly resolved near the gap regions are marked for each TM, i.e., $t_{2g}^{\uparrow,\downarrow}$ or $e_{g}^{\uparrow,\downarrow}$.
}
\end{figure}

Next, we studied the local magnetic states of TMs doped in SnTe, which are listed as $\mu$ in Table \ref{tab:mu}. The TMs from Ti to Ni show magnetic moments while Sc, Cu and Zn don't.
It is noted that the magnetic moments of the TMs doped in SnTe approximate to those on the isolated atoms in the vacuum. Among the magnetic TMs, Mn possesses the largest magnetic moment of 5 $\mu_B$.
In the octahedral crystal field, the $d$ orbitals split into two sets: the $d_{xy}$, $d_{xz}$, $d_{yz}$ orbitals with lower energy (known as $t_{2g}$) and the $d_{x^2-y^2}$, $d_{z^2}$ orbitals with higher energy (known as $e_{g}$). For the TM ion with the number of $d$ electrons between four and seven, they would show two possible electronic configurations called either high-spin/weak-field or low-spin/strong-field states respectively \cite{book_chemphy}, whose electronic configurations and magnetic moments are both given in Table. \ref{tab:mu}. In the rock-salt structure of SnTe, each ion is surrounded by six ions which have opposite charges forming a regular octahedron. It can be found that the TM ions with the number of $d$ electrons between four and seven all show high-spin states. On the other hand, the magnetic moments of Ti, Co and Ni are fractional, which may result from the hybridization with the bulk bands of SnTe.

The magnetic states of the $M_{\rm Sn}$ system could be clearly understood from the electronic structures of the system. Fig. \ref{fig:dos} displays the spin-polarized density of states (DOS) of TM-doped SnTe with the $d$ orbitals ($t_{2g}^{\uparrow,\downarrow}$ or $e_{g}^{\uparrow,\downarrow}$,) that could be clearly resolved near the gap region marked. Obviously, there is no spin splitting for Sc-, Cu- and Zn-doped SnTe, in consistence with their zero magnetic moments as shown in Table I.
From Ti- to Ni-doped SnTe, the spin splitting energy of $d$ orbitals is always stronger than the crystal splitting energy of the SnTe ligand, so the TM ion show high-spin states. There are localized $d$ orbitals in the gap regions for all magnetic systems except Mn-doped SnTe. In Mn-doped SnTe, the spin-up and the spin-down $d$ orbitals of Mn are all merged within the valence and conduction bands, respectively. The exclusion of defect levels near the gap region ensures an insulating bulk of SnTe and keeps the topology of SnTe.
When the partial filled $t_{2g}/e_g$ orbitals are close to the bulk bands of SnTe, the hybridization with the bulk bands may induce fractional filling of the $d$ orbitals, i.e., fractional magnetic moments, such as in the case of Ti, Co and Ni.
On the other hand, the 4$s$ orbitals of the TM atoms are generally unoccupied (not shown here), leaving the TM ions of formal +2 valence (equivalence substitution, as given in Table \ref{tab:mu}), except some special cases.
The $d$ orbitals of Sc are higher than the conduction band minimum (CBM), so the $d$ orbitals are unoccupied as well, leaving Sc with a valence state +3 and nonmagnetic. For Cr, the $e_{g}$ orbital is higher than the VBM, and the fifth $d$ electron occupies the valence band, turning the magnetic moment of Cr back to 4 $\mu$B, and be of +2 valence. While for Cu, the $d$ orbitals are all lower than the VBM of SnTe, ensuring its fully-filled $d$ orbitals kept and valence state of +1.

The hybridization between the $d$ orbitals of TMs and valence band of SnTe not only induces the changes of valence state and local magnetic moment of TMs, but contributes to the formation of the magnetic order. It has been reported that a carrier-concentration-induced ferromagnetic transitioin in Pb$_{1-x-y}$Sn$_y$Mn$_x$Te can be understood on the basis of the Ruderman-Kittel-Kasuya-Yosida (RKKY) interaction mechanism \cite{W.J.M.de-Jonge_semicondscitechnol_1990_carrier-MnSnTe} --- the coupling between the local magnetic moment and conduction electrons will influence the ferromagnetic interaction in the diluted magnetic semiconductors. The large magnetic moment of Mn and its hybridization with valence band of SnTe (shown in Fig. 4) 
effectively pave the way for the coupling between local magnetic moments via conduction electrons. In fact, from Fig. 4, other $M_{\rm {Sn}}$ ($M$ = Cr, Fe--Ni) systems have more hybridizations between the magnetic moments and the valence band of SnTe which are also expected to enable the RKKY-interaction-induced ferromagnetic order. However, it has never be observed in the latter systems \cite{Cr-clustering-in-SnTe,Cr-struct-transformation}, which may be attributed to the doping difficulty of the latter TM ions in SnTe, as was discussed earlier.
From the above analysis, Mn could introduce the magnetic order for the investigation of the theoretically-proposed QAHE in SnTe, and, moreover, it introduces no defect levels near the gap region. So in analogy with the prominent substitution of Cr or Fe in Bi$_{\rm 2}$Te$_{\rm 3}$-class TIs \cite{ZhangJianmin_PRB_2013_M-TI,cuizu_QAHE}, Mn is a promising magnetic dopant in SnTe.

\section{Conclusions}
In conclusion, using the first-principles calculations, we investigated the defect energetics and local magnetic states of the 3$d$ TMs dilutely doped in SnTe. Comparing to the case in Bi$_{\rm 2}$Te$_{\rm 3}$-class TIs --- V or Cr doping possesses a low formation energy providing the possibility of spontaneous doping, here the formation energies for the TMs doping in SnTe are comparatively high, which suggests that only a low doping concentration of TM is feasible experimentally. Besides, the formation energy is found positively correlated with the cohesive energy of the elemental TM bulk, and the TM dopants tend to stay in the neutral states when the Fermi level is located in the gap or below the VBM as in realistic SnTe.
In the dilute doping regime, all the TMs except Sc, Cu and Zn will introduce the local magnetic moments. Especially, the magnetic atoms with the $d$ electrons number between four and seven behave in high-spin states, indicating that the spin splitting energy of 3$d$ TM is stronger than the crystal splitting energy of the SnTe ligand. Among all the magnetic TMs, Mn has the largest magnetic moment and effective hybridization with the valence bands, enabling the formation of ferromagetic order. Moreover, Mn also shows relative low doping formation energy, and will not introduce defect level into the bulk gap of SnTe. These advantages make Mn the most promising dopant for SnTe. These results may provide a meaningful guide for the magnetic doping in TCIs and for the investigation of the QAHE with a large Chern number in SnTe.


We acknowledge the support of the Ministry of Science and Technology of China (Grant Nos. 2011CB921901 and 2011CB606405), and the National Natural Science Foundation of China (Grant No. 11334006).

\bibliographystyle{plain}

\end{document}